\documentclass[aip,letter,reprint]{revtex4-1}

\usepackage{graphicx}
\usepackage{subfig}

\begin{document}


\title{Ultrafast Linear Kinetic Inductive Photoresponse of {YBa$_2$Cu$_3$O$_{7-\delta}$} Meander-Line Structures by Photoimpedance Measurements} 

\author{Haig A. Atikian}
\email[Author to whom correspondence should be addressed. Electronic address: ]{haatikia@maxwell.uwaterloo.ca}
\affiliation{\footnotesize Electrical and Computer Engineering Department and the Institute for Quantum Computing, University of Waterloo, Waterloo, ON, N2L 3G1, Canada}
\author{Behnood G. Ghamsari}
\affiliation{\footnotesize Center for Nanophysics and Advanced
Materials, Department of Physics, University of Maryland, College
Park, MD, 20742-4111, USA}
\author{Steven M. Anlage}
\affiliation{\footnotesize Center for Nanophysics and Advanced
Materials, Department of Physics, University of Maryland, College
Park, MD, 20742-4111, USA}
\author{A. Hamed Majedi}
\affiliation{\footnotesize Electrical and Computer Engineering Department and the Institute for Quantum Computing, University of Waterloo, Waterloo, ON, N2L 3G1, Canada}

\date{\today}

\begin{abstract}
We report the experimental demonstration of the linear
kinetic-inductive photoresponse of thin-film
{YBa$_2$Cu$_3$O$_{7-\delta}$} (YBCO) meander-line structures, where
the photoresponse amplitude, full-width-half-maximum (FWHM), and
rise-time are bilinear in the incident optical power and bias
current. This bilinear behavior reveals a trade-off between
obtaining high responsivity and high speed photodetection. We also
report a rise-time as short as 29ps in our photoimpedance
measurements.
\end{abstract}


\maketitle

The interaction of light with superconducting samples is long known
to perturb superconductivity
\cite{Owen_PRL_72,Sai_PRL_74,Perrin_PRB_83,Frenkel_PRL_93}, which
can be used as a probing mechanism for optoelectronic
applications\cite{Testardi_PRB_71,Bluzer_PRB_91,Adam_TAS_99,Enomoto_JAP_86,Kwok_APL_89,Leung_APL_87,Brocklesby_APL_89,Forrester_TM_89,Hegmann_PRB_93}.
In general, photons of energy greater than the Cooper pair binding
energy ($2\Delta$) can initiate a chain of pair-breaking events
resulting in a deviation of the quasiparticle and pair densities
from their equilibrium values. Typically, these distributions depend
on temperature, optical power and wavelength, thermal boundary
conditions, and material properties such as electron-electron and
electron-phonon interactions times, electron density, coherence
length, penetration depth, and
geometry\cite{Bluzer_PRB_91,GapControl}. While determining the
spatial and temporal distribution of quasiparticles and pairs under
a time-varying optical illumination is a profound problem in
non-equilibrium superconductivity, many of the important concepts of
such an interaction for device applications can be captured by means
of a much simpler and more phenomenological approach, namely the
kinetic inductance model \cite{Hegmann_PRB_93}. Within the kinetic
inductance model the presence of the superconductive condensate, at
a macroscopic level, can be adequately modeled by an additional
inductive channel for charge transport.  The optically initiated
pair breaking mechanism, within this framework, should be
interpreted as the spatial and temporal variations of the kinetic
inductance and the normal resistance of the superconducting
specimen.

Many researchers have experimentally studied the kinetic inductive
photoresponse of superconducting thin films through photoimpedance
measurements \cite {Bluzer_PRB_91, Johnson_APL_91}. In
photoimpedance measurements, light induced changes in the microwave
impedance of the superconducting structure are measured by an
external high-frequency circuit. In its simplest form, the specimen
is externally biased with a dc current and connected to a fast
oscilloscope in series with a high bandwidth amplifier; absorption
of optical photons then changes the impedance of the sample and
produces a transient voltage response. A number of previous works
have reported photoimpedance measurements on different
superconductors mainly concluding that: 1) the resistive
photoresponse dominates at temperatures well below the critical
temperature (T$_c$), whereas the kinetic inductive response becomes
the main mechanism of photoresponse close to T$_c$; 2) In the
kinetic inductive regime the photoresponse could be very fast, with
a rise time as low as 50ps, and is mainly limited by the time
constants of the peripheral measuring apparatus; 3) the dependence
of the photoresponse amplitude varies nonlinearly with the incident
optical power.

The nonlinearity of the kinetic inductive photoresponse
intrinsically arises from the nonlinear dependence of the kinetic
inductance of a superconducting sample on the Cooper pair density.
Therefore, even though changes in the Cooper pair density, under
certain conditions, may vary linearly with the incident optical
power, the resultant variation of the kinetic inductance is
generally nonlinear. This point can be readily observed for a
thin-film sample\cite{Orlando}:
\begin{equation} \label{Lk}
L_k = \frac{m^\ast\ell}{n^\ast (q^\ast)^2A},
\end{equation}
where $L_k$ is the kinetic inductance, $\ell$ and $A$ are the length
and the cross section area of the sample, $m^\ast$ and $q^\ast$
respectively are the mass and charge of a Cooper pair, and $n^\ast$
is the density of Cooper pairs. Accordingly, the kinetic inductive
photoresponse approximately reads\cite{Adam_TAS_99}
\begin{equation}\label{LKID}
V_{ph} = \frac{d}{dt}(L_kI_0),
\end{equation}
where $I_0$ is the bias current. Nevertheless, we have theoretically
shown elsewhere\cite{Ghamsari_TAS_08, Ghamsari_Thesis} that if the
optical power and the bias current are far from their critical
values, and the temperature is not too close to T$_c$ the kinetic
inductive response can be linearized giving a frequency-dependent
voltage responsivity
\begin{equation} \label{Rv}
R_v(\omega) \equiv \frac{V_{ph}(\omega)}{P_o(\omega)} = \left(
\frac{\eta_Q\tau_QI_0}{2\Delta A\ell
n^\ast}\right)\left(\frac{j\omega L_{k0}R_{n0}}{j\omega
L_{k0}+R_{n0}}\right),
\end{equation}
where $P_o$ and $\omega$ are the incident optical power and
modulation frequency, $\eta_Q$ is the pair breaking efficiency,
$\tau_Q$ is the Cooper pairs recombination life time, and $R_{n0}$
and $L_{k0}$ are the equilibrium normal resistance and kinetic
inductance of the sample in the absence of illumination. This regime
of operation is particularly useful for optoelectronic device
applications such as photodetectors and optically tunable
microwave-photonic devices such as delay lines, resonators, and
filters where linear tunability is highly desirable \cite
{Atikian_MTT_2010}.

\begin{figure}
\centering{\subfloat[][]{\includegraphics[width =1.5in]{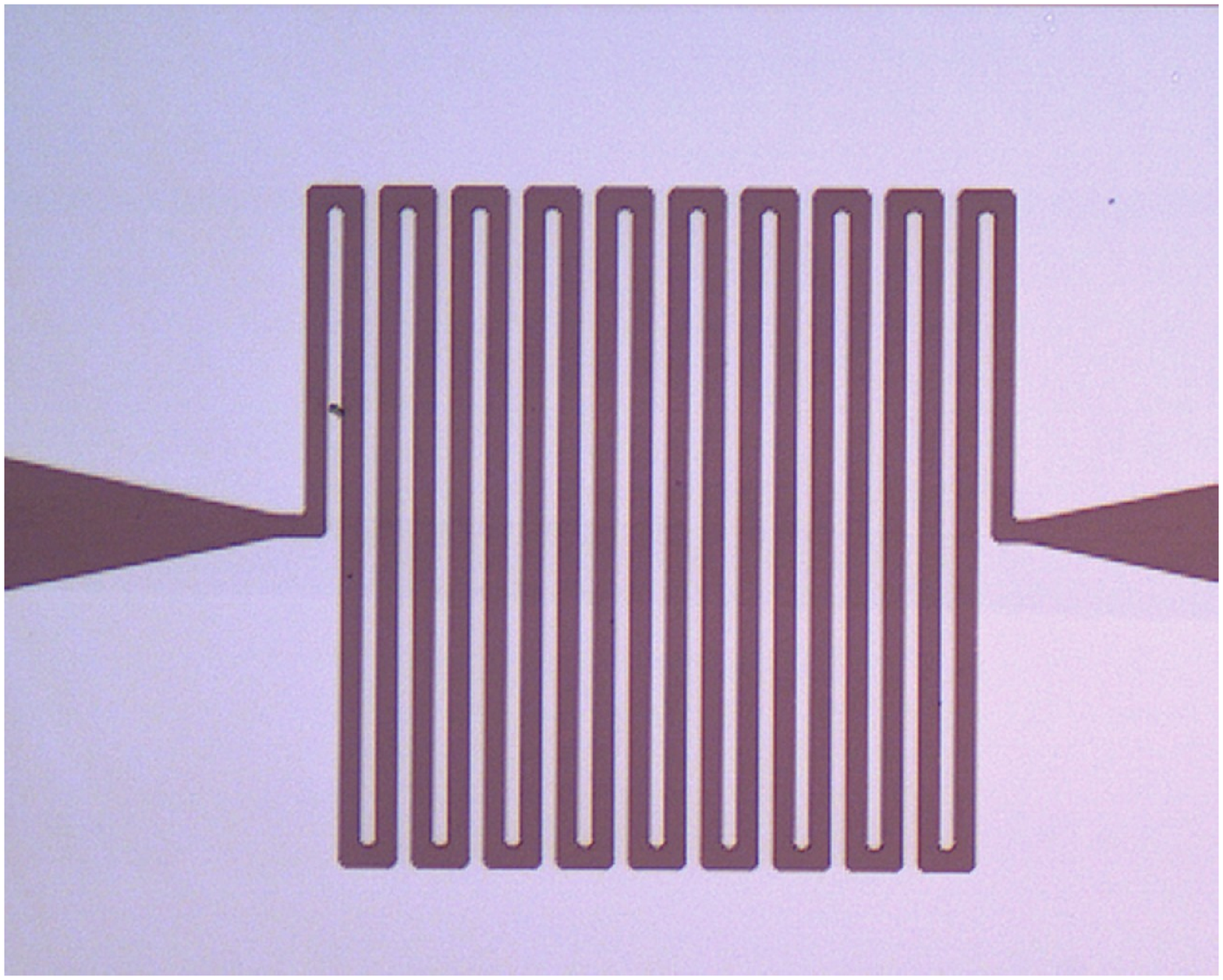}
\label{5um}}
\hspace{8pt}%
\subfloat[][]{\includegraphics[width=1.5in]{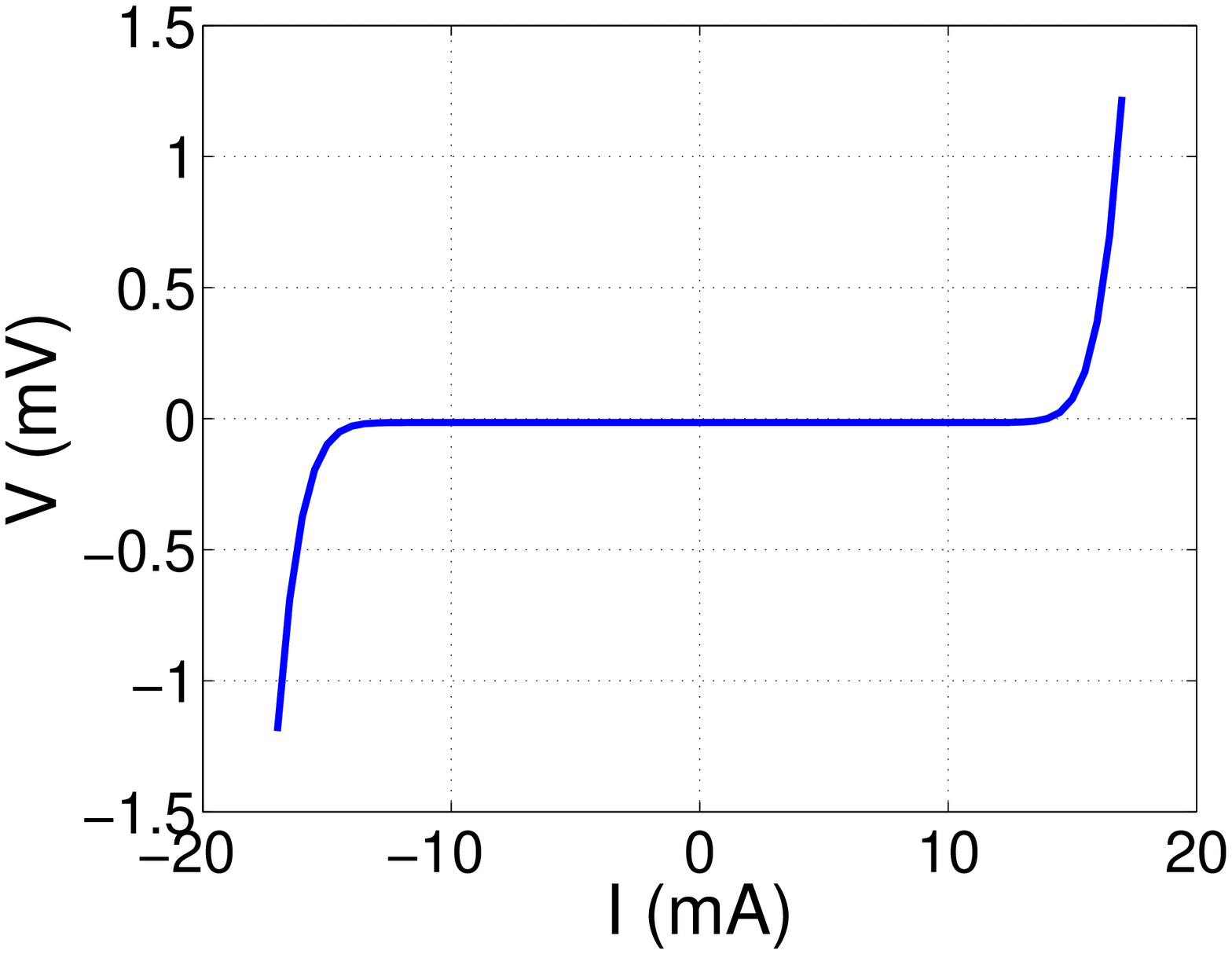}
\label{{IVML}}}} \caption{(a) Image of the 5$\mu m$ meander-line (b)
Measured current-voltage characteristics of the 5$\mu m$
meander-line at 77K.} \label{ML}
\end{figure}

\begin{figure}
\begin{center}
\includegraphics[width= 3.5 in]{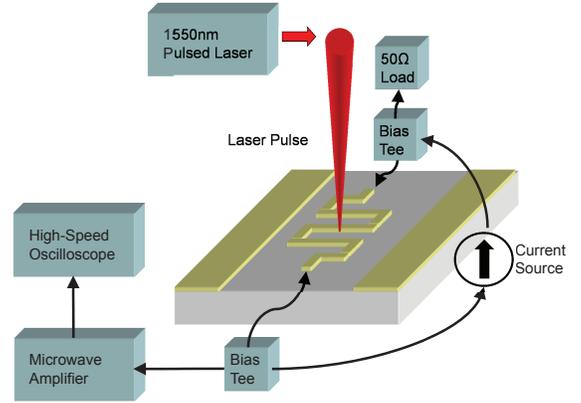}
\caption{Block diagram of the photoimpedance measurement setup of
the meander line structure.}\label{BlockDia}
\end{center}
\end{figure}

To serve as a detecting element, we have used a 100nm-thick YBCO
thin film (THEVA, Ismaning, Germany) meander line structure with
5-$\mu$m line widths and slots, covering an area of
176$\mu$m$\times$200$\mu$m.  The meander line structure is placed at
the midpoint of the center strip of a 50GHz-bandwidth 50$\Omega$
superconducting coplanar waveguide (CPW) transmission line. Figure
\ref{ML} shows the image of the meander line and the current-voltage
characteristics at 77K. The critical current is found to be 13mA,
and the bias current is selected to be below this value. The meander
line is externally dc biased through high bandwidth bias-tees. One
end of the CPW is terminated with a 50$\Omega$ load to suppress any
reflected signals. The other end is connected to a high-bandwidth
microwave amplifier with a gain of 28dB, followed by a fast
oscilloscope where the response to a train of 1550nm wavelength,
45ps-wide Gaussian optical pulses is measured. A block diagram of
the measurement setup is shown in Figure \ref{BlockDia}. More
details in regards to the photoimpedance experimental setup can be
found in \cite{Atikian_Thesis}.


\begin{figure}
\begin{center}
\includegraphics[width= 3.5 in]{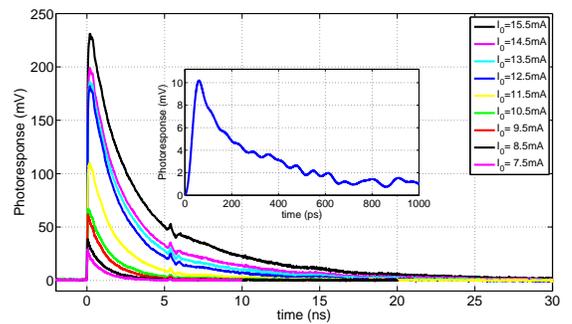}
\caption{Photoresponse waveforms for the 5$\mu m$ meander-line under
1.6mW of incident optical power with a varying bias at 77K and 28dB
amplification. (Inset) Photoresponse waveform under 1.2mW of
incident optical power and 7.5mA bias current with a 29ps
risetime.}\label{PR5}
\end{center}
\end{figure}

\begin{figure}
\centering{\subfloat[][]{\includegraphics[width
=1.5in]{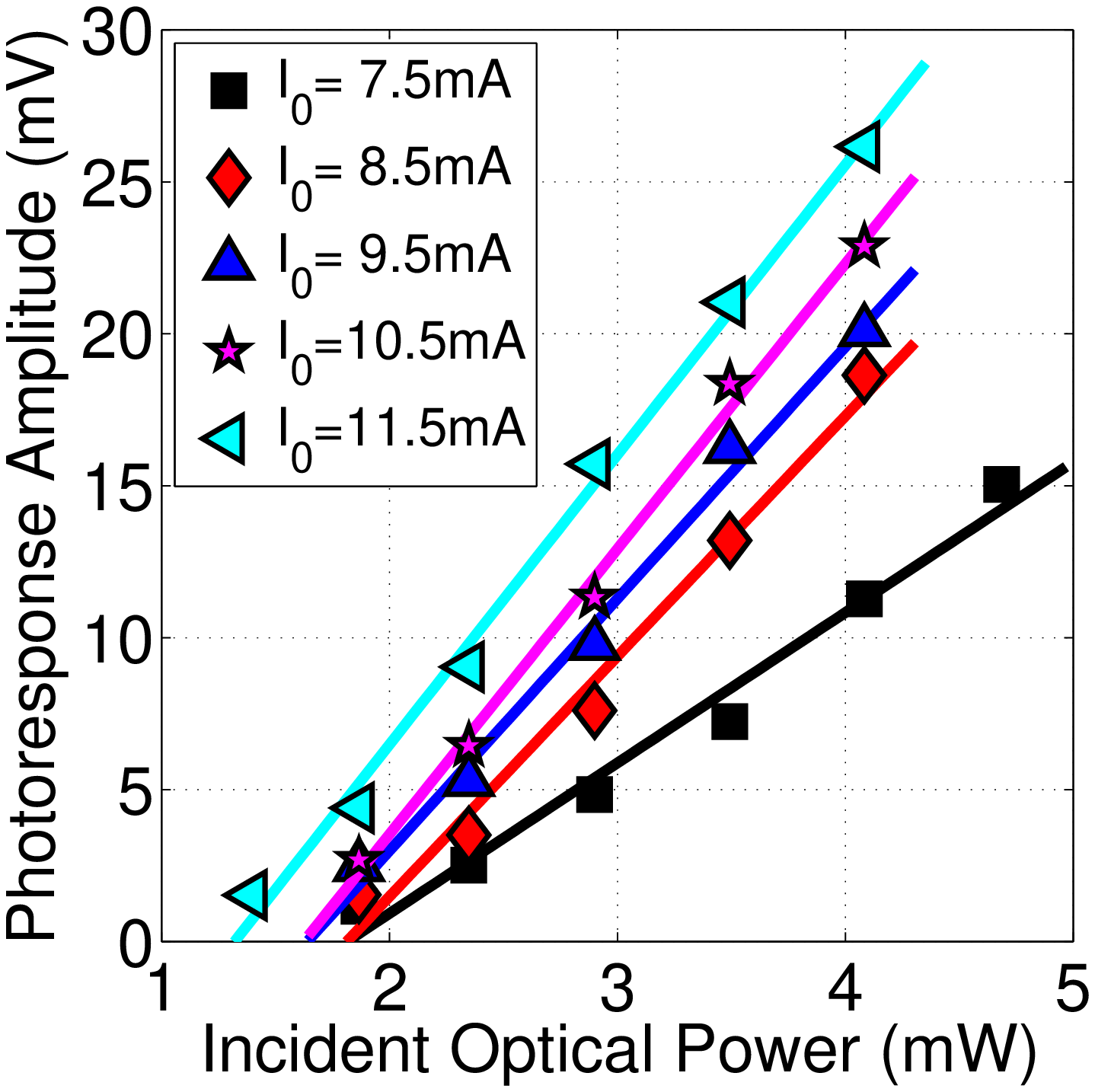} \label{AMPLITUDEa}}
\hspace{8pt}%
\subfloat[][]{\includegraphics[width=1.5in]{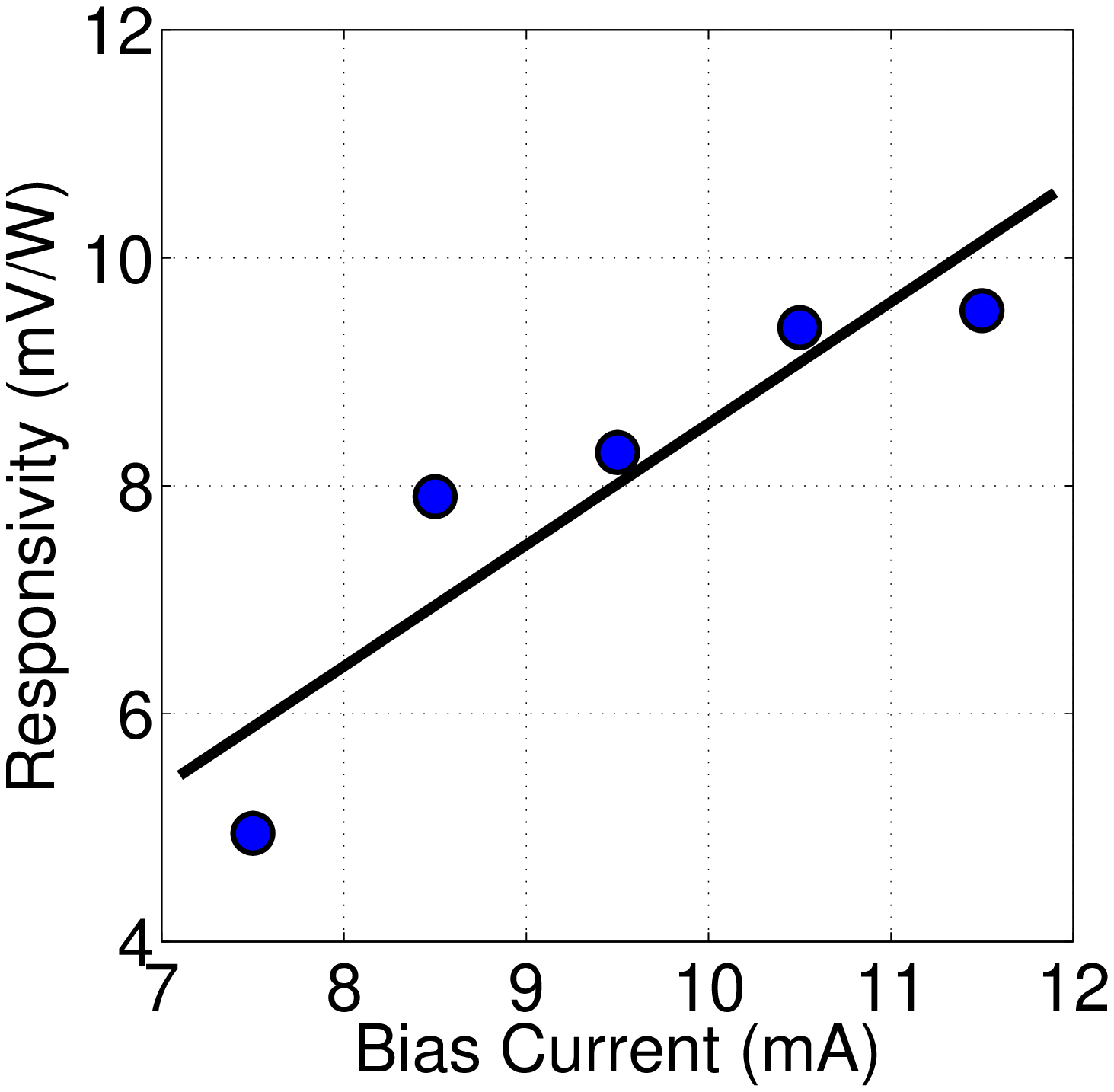}
\label{AMPLITUDEc}}} \caption{ (a) Photoresponse amplitude versus
incident optical power with a varying bias current for. (b)
Responsivity versus bias current.} \label{AMPLITUDE}
\end{figure}

Figure \ref{PR5} illustrates typical photoresponse waveforms for
different bias currents at an incident optical power of 1.6mW.  The
inset of Figure \ref{PR5} illustrates an operating point where we
have measured rise times as short as 29ps. Figure \ref{AMPLITUDEa}
shows that the photoresponse amplitudes of the detector, for fixed
bias currents, varies linearly with the incident optical power.
Moreover, Figure \ref{AMPLITUDEc} demonstrates that the responsivity
of the device has a linear dependence on the bias current. These two
observations confirm our previous theoretical prediction of the
linear kinetic inductive response represented by equation
(\ref{Rv}). This linear response sustains as long as the
perturbation in both kinetic inductance and normal resistance is
small. This also implies that the fractional change in both the
Cooper pair density and quasiparticles is small.  These values, in
general, depend on temperature, bias current, and average incident
optical power. The linear regime of operation for this device, at a
given temperature, is clearly illustrated by the range of current
and optical power values in Figure \ref{AMPLITUDE}.

\begin{figure}
\centering{\subfloat[][]{\includegraphics[width =1.5in]{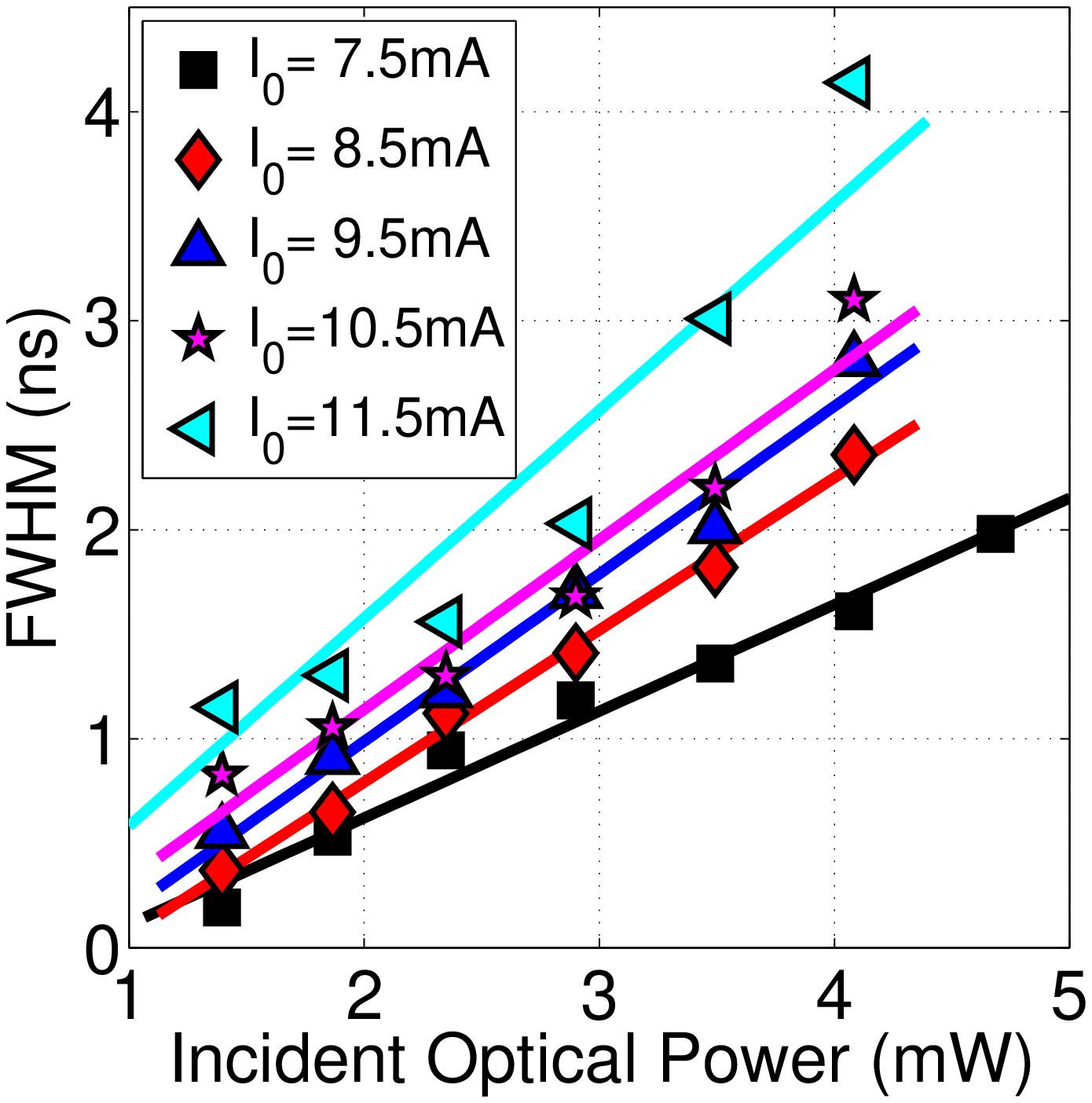}
\label{FWHMa}}
\hspace{8pt}%
\subfloat[][]{\includegraphics[width=1.5in]{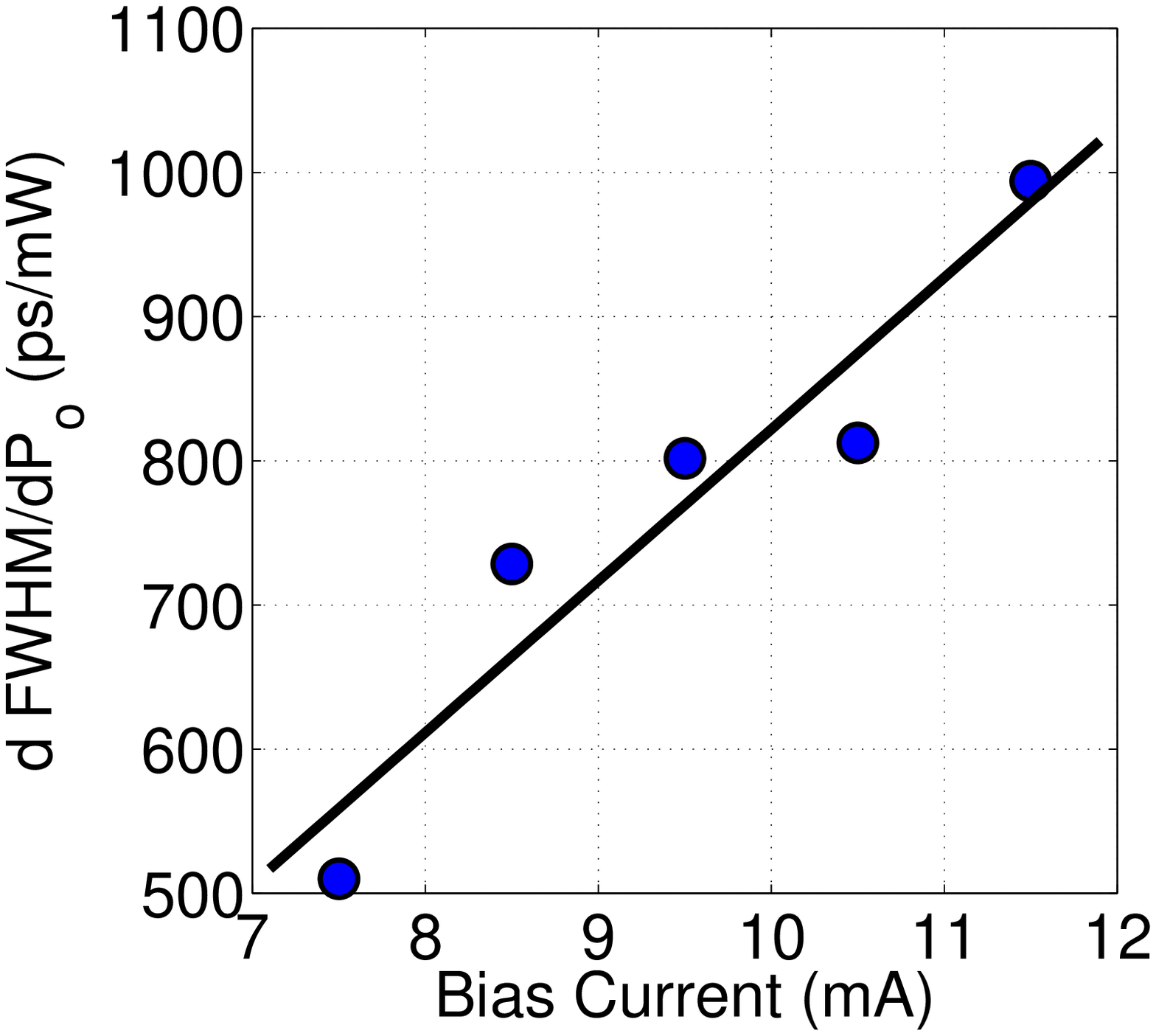}
\label{FWHMc}}} \caption{(a) FWHM of the photoresponse waveform
versus incident optical power with varying bias currents. (b) The
slope of the changes in FWHM with optical power as a function of
bias current.} \label{FWHM}
\end{figure}

The FWHM of the photoresponse is a measure of the photoinduced
disturbance in the detector, which according to (\ref{LKID}) equals
to $(\delta L_kI_0)$. The perturbation in the kinetic inductance,
$\delta L_k$, in the linear kinetic inductive regime, linearly
varies with optical power and is independent of the bias
current\cite{Ghamsari_Thesis}. Thus, the $(\delta L_kI_0)$ product,
and consequently the FWHM, should be bilinear in the optical power
and bias current which is clearly shown by Figure \ref{FWHM}. This
point readily reveals the trade off between obtaining high
responsivity and short photoresponse waveforms in linear kinetic
inductive detectors, because the former requires a higher bias
current whereas the latter demands a small bias current.

\begin{figure}
\centering{\subfloat[][]{\includegraphics[width =1.5in]{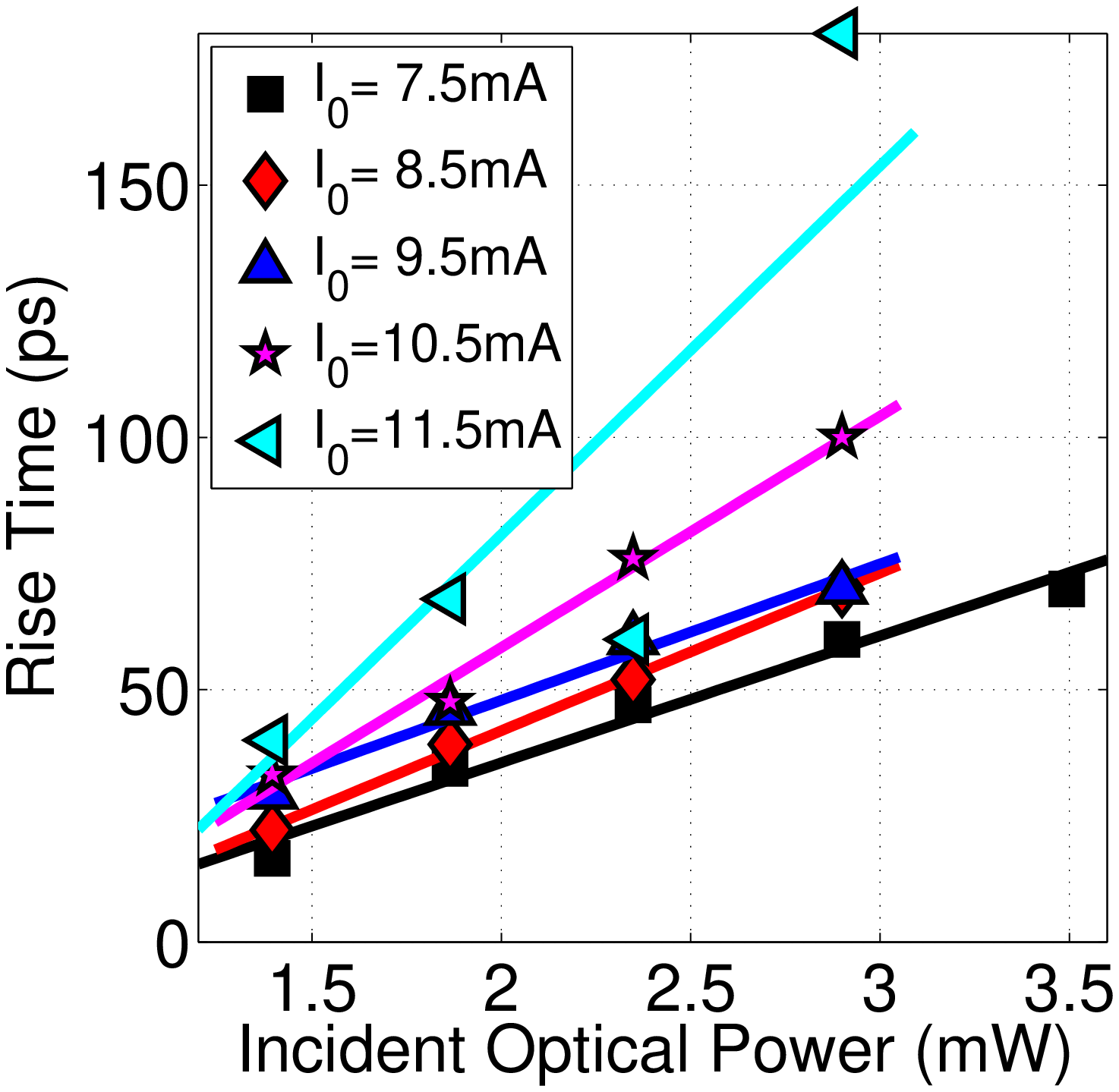}
\label{Tra}}
\hspace{8pt}%
\subfloat[][]{\includegraphics[width=1.5in]{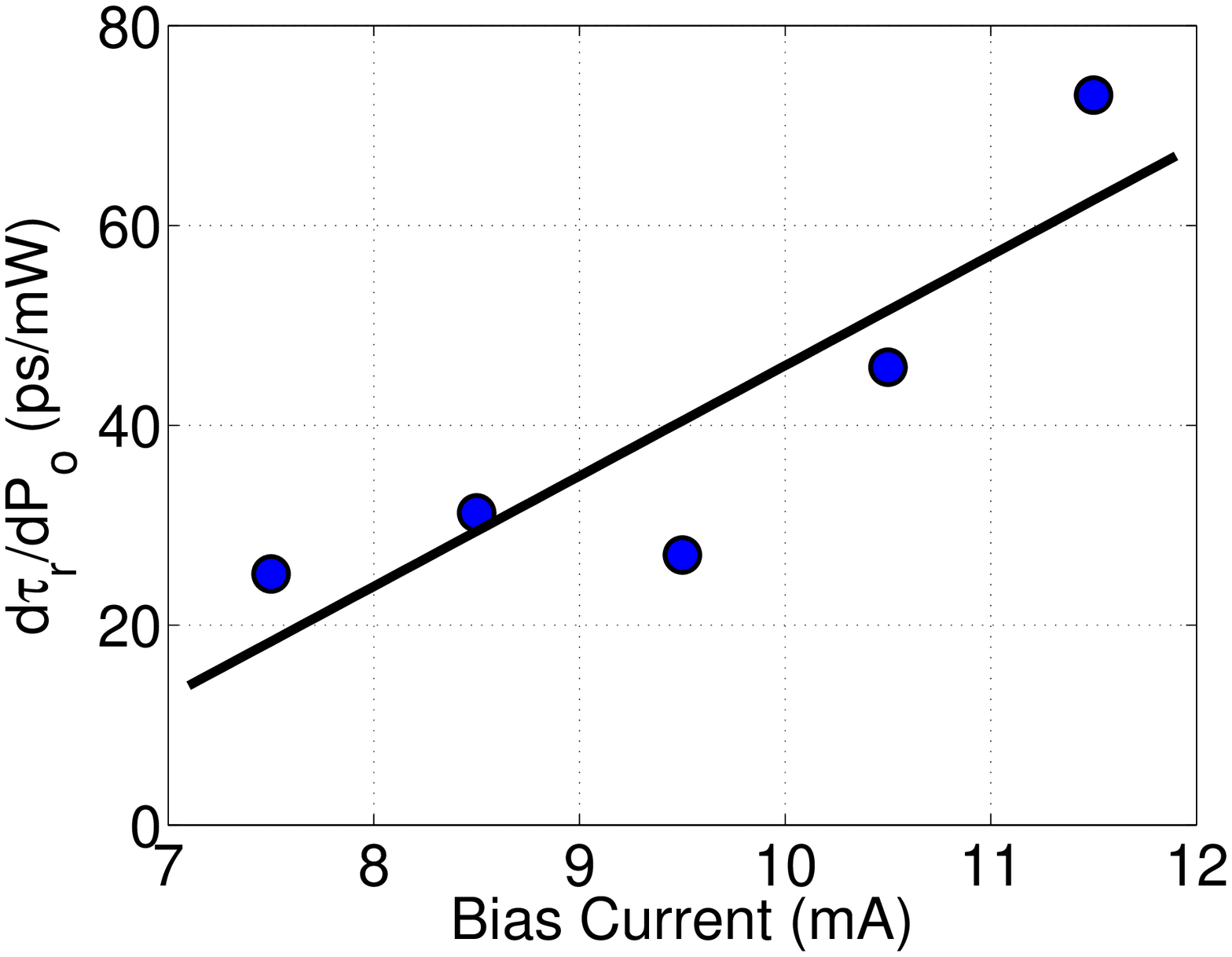} \label{Trc}}}
\caption{(a) Rise-time versus incident optical power with varying
bias currents (b) The slope of the changes in rise-time with optical
power as a function of bias current.} \label{Tr}
\end{figure}

In addition to amplitude and FWHM, the rise time of the
photoresponse waveform is also bilinear in optical power and bias
current, as illustrated by Figure \ref{Tr}. In terms of an
electrical circuit model, the detector acts like an RL circuit with
time varying $R$ and $L$, and a constant total current $I_0$
\cite{Ghamsari_Thesis}. In this scenario, the rise time depends on
the $(\delta L_kI_0)$, which was shown to be bilinear in current and
optical power. This point once again manifests itself as a trade-off
between speed and responsivity in the linear regime.  A set of
similar results was obtained for a meander line with 3$\mu$m line
and slot widths, for which the same trends as the presented device
was observed.

In conclusion, we have characterized the linear kinetic inductive
photoresponse of thin-film YBCO meander line structures, where the
photoresponse amplitude, FWHM, and rise time of the waveforms are
bilinear in optical power and bias current. For a given operating
point, we have been able to measure a short 29ps rise time in the
linear kinetic inductive regime.

We acknowledge support from OCE (Ontario Center of Excellence),
NSERC (Natural Sciences and Engineering Research Council of Canada),
and the ONR/Maryland Applied Electromagnetics Center Task D10
(contract No. N000140911190).

%


\begin{thebibliography}{20}%
\makeatletter
\providecommand \@ifxundefined [1]{%
 \@ifx{#1\undefined}
}%
\providecommand \@ifnum [1]{%
 \ifnum #1\expandafter \@firstoftwo
 \else \expandafter \@secondoftwo
 \fi
}%
\providecommand \@ifx [1]{%
 \ifx #1\expandafter \@firstoftwo
 \else \expandafter \@secondoftwo
 \fi
}%
\providecommand \natexlab [1]{#1}%
\providecommand \enquote  [1]{``#1''}%
\providecommand \bibnamefont  [1]{#1}%
\providecommand \bibfnamefont [1]{#1}%
\providecommand \citenamefont [1]{#1}%
\providecommand \href@noop [0]{\@secondoftwo}%
\providecommand \href [0]{\begingroup \@sanitize@url \@href}%
\providecommand \@href[1]{\@@startlink{#1}\@@href}%
\providecommand \@@href[1]{\endgroup#1\@@endlink}%
\providecommand \@sanitize@url [0]{\catcode `\\12\catcode
`\$12\catcode
  `\&12\catcode `\#12\catcode `\^12\catcode `\_12\catcode `\%12\relax}%
\providecommand \@@startlink[1]{}%
\providecommand \@@endlink[0]{}%
\providecommand \url  [0]{\begingroup\@sanitize@url \@url }%
\providecommand \@url [1]{\endgroup\@href {#1}{\urlprefix }}%
\providecommand \urlprefix  [0]{URL }%
\providecommand \Eprint [0]{\href }%
\providecommand \doibase [0]{http://dx.doi.org/}%
\providecommand \selectlanguage [0]{\@gobble}%
\providecommand \bibinfo  [0]{\@secondoftwo}%
\providecommand \bibfield  [0]{\@secondoftwo}%
\providecommand \translation [1]{[#1]}%
\providecommand \BibitemOpen [0]{}%
\providecommand \bibitemStop [0]{}%
\providecommand \bibitemNoStop [0]{.\EOS\space}%
\providecommand \EOS [0]{\spacefactor3000\relax}%
\providecommand \BibitemShut  [1]{\csname bibitem#1\endcsname}%
\let\auto@bib@innerbib\@empty
\bibitem [{\citenamefont {Owen}\ and\ \citenamefont
  {Scalapino}(1972)}]{Owen_PRL_72}%
  \BibitemOpen
  \bibfield  {author} {\bibinfo {author} {\bibfnamefont {C.~S.}\ \bibnamefont
  {Owen}}\ and\ \bibinfo {author} {\bibfnamefont {D.~J.}\ \bibnamefont
  {Scalapino}},\ }{\bibfield  {journal} {\bibinfo  {journal} {Phys.
  Rev. Lett.}\ }\textbf {\bibinfo {volume} {28}},\ \bibinfo {pages}
  {1559--1561} (\bibinfo {year} {1972})}\BibitemShut {NoStop}%
\bibitem [{\citenamefont {Sai-Halasz}\ \emph {et~al.}(1974)\citenamefont
  {Sai-Halasz}, \citenamefont {Chi}, \citenamefont {Denenstein},\ and\
  \citenamefont {Langenberg}}]{Sai_PRL_74}%
  \BibitemOpen
  \bibfield  {author} {\bibinfo {author} {\bibfnamefont {G.~A.}\ \bibnamefont
  {Sai-Halasz}}, \bibinfo {author} {\bibfnamefont {C.~C.}\ \bibnamefont {Chi}},
  \bibinfo {author} {\bibfnamefont {A.}~\bibnamefont {Denenstein}}, \ and\
  \bibinfo {author} {\bibfnamefont {D.~N.}\ \bibnamefont {Langenberg}},\
  }{\bibfield
  {journal} {\bibinfo  {journal} {Phys. Rev. Lett.}\ }\textbf {\bibinfo
  {volume} {33}},\ \bibinfo {pages} {215--219} (\bibinfo {year}
  {1974})}\BibitemShut {NoStop}%
\bibitem [{\citenamefont {Perrin}\ and\ \citenamefont
  {Vanneste}(1983)}]{Perrin_PRB_83}%
  \BibitemOpen
  \bibfield  {author} {\bibinfo {author} {\bibfnamefont {N.}~\bibnamefont
  {Perrin}}\ and\ \bibinfo {author} {\bibfnamefont {C.}~\bibnamefont
  {Vanneste}},\ }
  {\bibfield  {journal} {\bibinfo  {journal} {Phys. Rev. B}\ }\textbf {\bibinfo
  {volume} {28}},\ \bibinfo {pages} {5150--5159} (\bibinfo {year}
  {1983})}\BibitemShut {NoStop}%
\bibitem [{\citenamefont {Frenkel}(1993)}]{Frenkel_PRL_93}%
  \BibitemOpen
  \bibfield  {author} {\bibinfo {author} {\bibfnamefont {A.}~\bibnamefont
  {Frenkel}},\ }{\bibfield  {journal} {\bibinfo  {journal} {Phys. Rev. B}\
  }\textbf {\bibinfo {volume} {48}},\ \bibinfo {pages} {9717--9725} (\bibinfo
  {year} {1993})}\BibitemShut {NoStop}%
\bibitem [{\citenamefont {Testardi}(1971)}]{Testardi_PRB_71}%
  \BibitemOpen
  \bibfield  {author} {\bibinfo {author} {\bibfnamefont {L.~R.}\ \bibnamefont
  {Testardi}},\ }{\bibfield  {journal}
  {\bibinfo  {journal} {Phys. Rev. B}\ }\textbf {\bibinfo {volume} {4}},\
  \bibinfo {pages} {2189--2196} (\bibinfo {year} {1971})}\BibitemShut {NoStop}%
\bibitem [{\citenamefont {Bluzer}(1991)}]{Bluzer_PRB_91}%
  \BibitemOpen
  \bibfield  {author} {\bibinfo {author} {\bibfnamefont {N.}~\bibnamefont
  {Bluzer}},\ }{\bibfield  {journal} {\bibinfo
  {journal} {Phys. Rev. B}\ }\textbf {\bibinfo {volume} {44}},\ \bibinfo
  {pages} {10222--10233} (\bibinfo {year} {1991})}\BibitemShut {NoStop}%
\bibitem [{\citenamefont {Adam}\ \emph {et~al.}(1999)\citenamefont {Adam},
  \citenamefont {Currie}, \citenamefont {Sobolewski}, \citenamefont {Harnack},\
  and\ \citenamefont {Darula}}]{Adam_TAS_99}%
  \BibitemOpen
  \bibfield  {author} {\bibinfo {author} {\bibfnamefont {R.}~\bibnamefont
  {Adam}}, \bibinfo {author} {\bibfnamefont {M.}~\bibnamefont {Currie}},
  \bibinfo {author} {\bibfnamefont {R.}~\bibnamefont {Sobolewski}}, \bibinfo
  {author} {\bibfnamefont {O.}~\bibnamefont {Harnack}}, \ and\ \bibinfo
  {author} {\bibfnamefont {M.}~\bibnamefont {Darula}},\ }{\bibfield  {journal} {\bibinfo  {journal} {IEEE Trans. Appl.
  Supercond.}\ }\textbf {\bibinfo {volume} {9}},\ \bibinfo {pages} {4091--4094}
  (\bibinfo {year} {1999})}\BibitemShut {NoStop}%
\bibitem [{\citenamefont {Enomoto}\ and\ \citenamefont
  {Murakami}(1986)}]{Enomoto_JAP_86}%
  \BibitemOpen
  \bibfield  {author} {\bibinfo {author} {\bibfnamefont {Y.}~\bibnamefont
  {Enomoto}}\ and\ \bibinfo {author} {\bibfnamefont {T.}~\bibnamefont
  {Murakami}},\ }
  {\bibfield  {journal} {\bibinfo  {journal} {J. Appl. Phys.}\ }\textbf
  {\bibinfo {volume} {59}},\ \bibinfo {pages} {3807—3814} (\bibinfo {year}
  {1986})}\BibitemShut {NoStop}%
\bibitem [{\citenamefont {Kwok}, \citenamefont {Zheng},\ and\ \citenamefont
  {Ying}(1989)}]{Kwok_APL_89}%
  \BibitemOpen
  \bibfield  {author} {\bibinfo {author} {\bibfnamefont {H.~S.}\ \bibnamefont
  {Kwok}}, \bibinfo {author} {\bibfnamefont {J.~P.}\ \bibnamefont {Zheng}}, \
  and\ \bibinfo {author} {\bibfnamefont {Q.~Y.}\ \bibnamefont {Ying}},\
  }{\bibfield  {journal} {\bibinfo
  {journal} {Appl. Phys. Lett.}\ }\textbf {\bibinfo {volume} {54}},\ \bibinfo
  {pages} {2473--2475} (\bibinfo {year} {1989})}\BibitemShut {NoStop}%
\bibitem [{\citenamefont {Leung}\ \emph {et~al.}(1987)\citenamefont {Leung},
  \citenamefont {Broussard}, \citenamefont {Claassen}, \citenamefont {Usofsky},
  \citenamefont {Wolf},\ and\ \citenamefont {Strum}}]{Leung_APL_87}%
  \BibitemOpen
  \bibfield  {author} {\bibinfo {author} {\bibfnamefont {M.}~\bibnamefont
  {Leung}}, \bibinfo {author} {\bibfnamefont {P.~R.}\ \bibnamefont
  {Broussard}}, \bibinfo {author} {\bibfnamefont {J.~H.}\ \bibnamefont
  {Claassen}}, \bibinfo {author} {\bibfnamefont {M.}~\bibnamefont {Usofsky}},
  \bibinfo {author} {\bibfnamefont {S.~A.}\ \bibnamefont {Wolf}}, \ and\
  \bibinfo {author} {\bibfnamefont {U.}~\bibnamefont {Strum}},\ }
  {\bibfield  {journal} {\bibinfo  {journal} {Appl. Phys. Lett.}\ }\textbf
  {\bibinfo {volume} {51}},\ \bibinfo {pages} {2046--2049} (\bibinfo {year}
  {1987})}\BibitemShut {NoStop}%
\bibitem [{\citenamefont {Brocklesby}\ \emph {et~al.}(1989)\citenamefont
  {Brocklesby}, \citenamefont {Monroe}, \citenamefont {Levi}, \citenamefont
  {Hong}, \citenamefont {Liou}, \citenamefont {Kwo}, \citenamefont {Rice},
  \citenamefont {Mankiewich},\ and\ \citenamefont
  {Howard}}]{Brocklesby_APL_89}%
  \BibitemOpen
  \bibfield  {author} {\bibinfo {author} {\bibfnamefont {W.~S.}\ \bibnamefont
  {Brocklesby}}, \bibinfo {author} {\bibfnamefont {D.}~\bibnamefont {Monroe}},
  \bibinfo {author} {\bibfnamefont {A.~F.~J.}\ \bibnamefont {Levi}}, \bibinfo
  {author} {\bibfnamefont {M.}~\bibnamefont {Hong}}, \bibinfo {author}
  {\bibfnamefont {S.~H.}\ \bibnamefont {Liou}}, \bibinfo {author}
  {\bibfnamefont {J.}~\bibnamefont {Kwo}}, \bibinfo {author} {\bibfnamefont
  {C.~E.}\ \bibnamefont {Rice}}, \bibinfo {author} {\bibfnamefont {P.~M.}\
  \bibnamefont {Mankiewich}}, \ and\ \bibinfo {author} {\bibfnamefont {R.~E.}\
  \bibnamefont {Howard}},\ }{\bibfield  {journal} {\bibinfo  {journal} {Appl.
  Phys. Lett.}\ }\textbf {\bibinfo {volume} {54}},\ \bibinfo {pages}
  {1175—1177} (\bibinfo {year} {1989})}\BibitemShut {NoStop}%
\bibitem [{\citenamefont {Forrester}\ \emph {et~al.}(1989)\citenamefont
  {Forrester}, \citenamefont {Gottlieb}, \citenamefont {Gavaler},\ and\
  \citenamefont {Braginski}}]{Forrester_TM_89}%
  \BibitemOpen
  \bibfield  {author} {\bibinfo {author} {\bibfnamefont {M.~G.}\ \bibnamefont
  {Forrester}}, \bibinfo {author} {\bibfnamefont {M.}~\bibnamefont {Gottlieb}},
  \bibinfo {author} {\bibfnamefont {J.~R.}\ \bibnamefont {Gavaler}}, \ and\
  \bibinfo {author} {\bibfnamefont {A.~I.}\ \bibnamefont {Braginski}},\
  }{\bibfield  {journal} {\bibinfo
  {journal} {IEEE Trans. Magnet.}\ }\textbf {\bibinfo {volume} {25}},\ \bibinfo
  {pages} {1327--1330} (\bibinfo {year} {1989})}\BibitemShut {NoStop}%
\bibitem [{\citenamefont {Hegmann}\ and\ \citenamefont
  {Preston}(1993)}]{Hegmann_PRB_93}%
  \BibitemOpen
  \bibfield  {author} {\bibinfo {author} {\bibfnamefont {F.~A.}\ \bibnamefont
  {Hegmann}}\ and\ \bibinfo {author} {\bibfnamefont {J.~S.}\ \bibnamefont
  {Preston}},\ }{\bibfield  {journal} {\bibinfo  {journal} {Phys. Rev. B}\
  }\textbf {\bibinfo {volume} {48}},\ \bibinfo {pages} {16023--16039} (\bibinfo
  {year} {1993})}\BibitemShut {NoStop}%
\bibitem [{\citenamefont {Pals}\ \emph {et~al.}(1982)\citenamefont {Pals},
  \citenamefont {Weiss}, \citenamefont {{van Atfekum}}, \citenamefont
  {Horstman},\ and\ \citenamefont {Wolter}}]{GapControl}%
  \BibitemOpen
  \bibfield  {author} {\bibinfo {author} {\bibfnamefont {J.~A.}\ \bibnamefont
  {Pals}}, \bibinfo {author} {\bibfnamefont {K.}~\bibnamefont {Weiss}},
  \bibinfo {author} {\bibfnamefont {P.~M. T.~M.}\ \bibnamefont {{van
  Atfekum}}}, \bibinfo {author} {\bibfnamefont {R.~E.}\ \bibnamefont
  {Horstman}}, \ and\ \bibinfo {author} {\bibfnamefont {J.}~\bibnamefont
  {Wolter}},\ }{\bibfield
  {journal} {\bibinfo  {journal} {Phys. Rep.}\ }\textbf {\bibinfo {volume}
  {89}},\ \bibinfo {pages} {323--390} (\bibinfo {year} {1982})}\BibitemShut
  {NoStop}%
\bibitem [{\citenamefont {Johnson}(1991)}]{Johnson_APL_91}%
  \BibitemOpen
  \bibfield  {author} {\bibinfo {author} {\bibfnamefont {M.}~\bibnamefont
  {Johnson}},\ }{\bibfield  {journal} {\bibinfo  {journal}
  {Appl. Phys. Lett.}\ }\textbf {\bibinfo {volume} {59}},\ \bibinfo {pages}
  {{1371--1373}} (\bibinfo {year} {1991})}\BibitemShut {NoStop}%
\bibitem [{\citenamefont {Orlando}\ and\ \citenamefont
  {Delin}(1991)}]{Orlando}%
  \BibitemOpen
  \bibfield  {author} {\bibinfo {author} {\bibfnamefont {T.~P.}\ \bibnamefont
  {Orlando}}\ and\ \bibinfo {author} {\bibfnamefont {K.~A.}\ \bibnamefont
  {Delin}},\ }\href@noop {} \ (\bibinfo  {publisher} {Addison-Wesley},\ \bibinfo
  {address} {Massachusetts},\ \bibinfo {year} {1991})\BibitemShut {NoStop}%
\bibitem [{\citenamefont {Ghamsari}\ and\ \citenamefont
  {Majedi}(2008)}]{Ghamsari_TAS_08}%
  \BibitemOpen
  \bibfield  {author} {\bibinfo {author} {\bibfnamefont {B.~G.}\ \bibnamefont
  {Ghamsari}}\ and\ \bibinfo {author} {\bibfnamefont {A.~H.}\ \bibnamefont
  {Majedi}},\ } {\bibfield
  {journal} {\bibinfo  {journal} {Trans. Appl. Supercond.}\ }\textbf {\bibinfo
  {volume} {18}},\ \bibinfo {pages} {1761--1768} (\bibinfo {year}
  {2008})}\BibitemShut {NoStop}%
\bibitem [{\citenamefont {Ghamsari}(2010)}]{Ghamsari_Thesis}%
  \BibitemOpen
  \bibfield  {author} {\bibinfo {author} {\bibfnamefont {B.~G.}\ \bibnamefont
  {Ghamsari}},\ }{Ph.D. thesis},\ \bibinfo  {school}
  {University of Waterloo} (\bibinfo {year} {2010})\BibitemShut {NoStop}%
\bibitem [{\citenamefont {Atikian}, \citenamefont {Ghamsari},\ and\
  \citenamefont {Majedi}(2010)}]{Atikian_MTT_2010}%
  \BibitemOpen
  \bibfield  {author} {\bibinfo {author} {\bibfnamefont {H.~A.}\ \bibnamefont
  {Atikian}}, \bibinfo {author} {\bibfnamefont {B.~G.}\ \bibnamefont
  {Ghamsari}}, \ and\ \bibinfo {author} {\bibfnamefont {A.~H.}\ \bibnamefont
  {Majedi}},\ } {\bibfield  {journal}
  {\bibinfo  {journal} {IEEE Trans. Microwave Theory Tech.}\ }\textbf {\bibinfo
  {volume} {58}},\ \bibinfo {pages} {3320--3326} (\bibinfo {year}
  {2010})}\BibitemShut {NoStop}%
\bibitem [{\citenamefont {Atikian}(2009)}]{Atikian_Thesis}%
  \BibitemOpen
  \bibfield  {author} {\bibinfo {author} {\bibfnamefont {H.~A.}\ \bibnamefont
  {Atikian}},\ }{Master's thesis},\ \bibinfo  {school} {University of Waterloo} (\bibinfo
  {year} {2009})\BibitemShut {NoStop}%
\end{thebibliography}
\end{document}